# Investigating Voice as a Biomarker for leucine-rich repeat kinase 2-Associated Parkinson's Disease

Siddharth Arora, PhD,[1†] Naomi P. Visanji, PhD,[2†] Tiago A. Mestre, MD, MSc,[3] Athanasios Tsanas, PhD,[4] Amaal AlDakheel, MD,[2] Barbara S. Connolly, MD,[5] Carmen Gasca-Salas, MD, PhD,[2] Drew S. Kern, MD, MSc,[6] Jennifer Jain, MD,[2] Elizabeth J. Slow, MD, PhD,[2] Achinoam Faust-Socher, MD,[2] Anthony E. Lang, MD,[2] Max A. Little, PhD,[7,8] and Connie Marras, MD, PhD[2]

†denotes equal contribution

[1]Somerville College, University of Oxford, Oxford, UK

[2]The Edmond J. Safra Program in Parkinson's Disease and the Morton and Gloria Shulman Movement Disorders Centre and, Toronto Western Hospital, Toronto, Ontario, Canada

[3]Parkinson's Disease and Movement Disorders Center, Division of Neurology, Department of Medicine, The Ottawa Hospital Research Institute, University of Ottawa Brain and Mind Institute, Ottawa, Canada

[4]Usher Institute of Population Health Sciences and Informatics, University of Edinburgh, Edinburgh, UK

[5]Division of Neurology, Department of Medicine, Hamilton Health Sciences, McMaster University, Hamilton, Ontario, Canada

[6]Movement Disorders Center, Department of Neurology, University of Colorado, Anschutz Medical Campus, Aurora, Colorado, USA

[7]Engineering and Applied Science, Aston University, Birmingham, UK

[8]Media Lab, Massachusetts Institute of Technology, Cambridge, MA, USA

**Corresponding Author:**
Name: Dr. Connie Marras
Email: Connie.Marras@uhnresearch.ca
Phone: +1-416-603-6422
Permanent address: 84 Queen's Park, Toronto, Ontario M5S 2C5, Canada

**Abstract**

We investigate the potential association between leucine-rich repeat kinase 2 (*LRRK2*) mutations and voice. Sustained phonations ('aaah' sounds) were recorded from 7 individuals with *LRRK2*-associated Parkinson's disease (PD), 17 participants with idiopathic PD (iPD), 20 non-manifesting *LRRK2*-mutation carriers, 25 related non-carriers, and 26 controls. In distinguishing *LRRK2*-associated PD and iPD, the mean sensitivity was 95.4% (SD 17.8%) and mean specificity was 89.6% (SD 26.5%). Voice features for non-manifesting carriers, related non-carriers, and controls were much less discriminatory. Vocal deficits in *LRRK2*-associated PD may be different than those in iPD. These preliminary results warrant longitudinal analyses and replication in larger cohorts.

**Introduction**

Voice impairment may be one of the earliest motor indicators of idiopathic Parkinson's disease (iPD) [1], and is typically characterized by breathiness, roughness, reduced loudness, and vocal tremor [1-4]. It is estimated that between 70 and 90% of people with PD (PWP) experience vocal impairment [2, 5, 6], and nearly one-third of PWP report voice-related problems as one of their main disease-related limitations [6].

Past work has demonstrated that objective measures of vocal impairment can be used to distinguish participants with iPD from controls with a high accuracy (mean sensitivity and mean specificity > 90%) [7-15]. The extent of vocal dysfunction has also been shown to be associated with disease severity [3, 16, 17]. Moreover, for symptom monitoring, voice-based measures have been used to accurately replicate both the motor and total Unified Parkinson's Disease Rating Scale (UPDRS) assessment (within 2 points from the clinicians' estimate) [18]. Recently, abnormalities in speech production have also been reported in participants with idiopathic rapid eye movement (REM) sleep behaviour disorder [19, 20]. These findings encourage further investigation of voice analysis as a reliable, non-invasive, and scalable tool that may also identify prodromal PD.

Leucine-rich repeat kinase 2 (*LRRK2*) mutations are the most common cause of genetically-determined PD [21]. The opportunity to intervene with disease-modifying therapy early in the neurodegenerative process makes identifying the prodromal state important. To investigate the presence of voice abnormalities in populations at increased risk to develop PD, in this study we analyse voice-based measures in multiplex families carrying a mutation in the gene for *LRRK2*. The goals of this pilot study were thus twofold. First, we aimed to determine if voice can be used to discriminate participants with *LRRK2*- associated PD from idiopathic PD. Second, we examined if there are any differences in voice between non-manifesting carriers of a *LRRK2* mutation when compared to related non-manifesting non-mutation carriers and unrelated healthy controls.

**Methods**

*Study participants*

Probands with *LRRK2* mutations were identified at Toronto Western Hospital and all available blood relatives were invited to participate. iPD patients and healthy individuals (devoid of any neurologic disease or family history of PD) were recruited at Toronto Western Hospital. iPD was defined as individuals with PD, according to clinical diagnosis by a movement disorder specialist, in the absence of a family history of the disease in a first or second-degree relative. Seven participants with *LRRK2*-PD (p.P.G2019S (5) or L1795F (2)), 17 participants with iPD, 20 non-manifesting carriers of *LRRK2* mutations (p.G2019S (18), L1795F (2)), 25 related non-manifesting non-carriers, and 26 healthy controls were recruited. The presence or absence of a *LRRK2* mutation was evaluated in all participants as described previously [22]. In the non-manifesting carrier group, the likelihood of prodromal disease being present was determined [23]. The study was approved by the University Health Network Research Ethics Board and informed consent was obtained from all participants.

*Data acquisition*

We obtained two audio recordings of sustained vocal phonation from each participant during a study visit at the Toronto Western Hospital using a USB powered microphone (Logitech, model 980186-0403) positioned on a stable surface ~2 inches from the participant's mouth. Recordings were collected using Audacity software (Version 2.0.3) in a quiet room. Participants with iPD were evaluated in the ON medication state. Each participant was instructed to "Take a deep breath and then let out a single "aaah" sound for as long as you can." Each recording was sampled at 44.1 kHz and stored as a de-identified digital audio file (.wav format).

*Data processing*

Identification of the longest usable segment of sustained phonation for each recording was performed manually. Recordings were discarded from the analysis if they were too noisy or if the phonation duration was shorter than two seconds. For each recording, we extracted 292 summary measures (also referred to as *features* or *dysphonia measures*) that have been used for analysing voice, including in PD [10, 14, 18, 24]. Details regarding these features are provided in the Supplementary section.

*Statistical analysis*

We identified 3 pairwise comparisons of interest: (1) *LRRK2*-PD versus iPD, (2) Non-manifesting carriers versus related non-manifesting non-carriers, and, (3) Non-manifesting carriers versus healthy controls. For each pairwise comparison, salient features were identified using the following 5 feature selection algorithms that help enhance the explanatory power of the analysis by removing redundant and less informative features [25-29]. Each of the 5 feature selection algorithms provided a unique set of feature ranking. To obtain a single ranking of the most salient features to be used for group comparison, we used a majority voting scheme. Pairwise comparisons were performed using a highly nonlinear statistical machine learning algorithm (random forests), used to separate generic feature data into several different classes [30]. Discrimination accuracy was evaluated using a 10-fold cross-validation (CV) scheme (with 100 repetitions for statistical confidence). This scheme helps assess generalizability of the discrimination results to similar, but previously unseen data, and has been used in previous studies on voice analysis in PD [10, 14, 18]. Data was balanced in each cross-validation repetition to eliminate differences in group sample size. The statistical significance level was set to *p*=0.05. Statistical analysis of the voice recordings was performed using the Matlab® software (version 2016b). Details regarding statistical analysis focussing on feature extraction, feature selection, and validation are provided in the Supplementary section.

**Results**

*LRRK2-PD vs idiopathic PD*

On average, *LRRK2*-PD participants were older and had a longer disease duration compared to participants with iPD (Table 1). However, UPDRSIII (motor UPDRS) between the two groups was similar. Two recordings were collected from each participant, however, 3 *LRRK2*-PD and 2 iPD voice recordings were discarded as they were too noisy for reliable computation of features. Accuracies to distinguish *LRRK2* PD from iPD are reported in Table 1 and were computed using the 10 most salient features. Including more features in the classifier (random forest) improved the discrimination accuracy only marginally (Supplementary Figure 2). In discriminating participants using 11 *LRRK2*-PD (n = 7 individuals) and 32 iPD voice recordings (n = 17 individuals), the mean sensitivity was 95.4% (Standard Deviation (SD) 17.8%) and mean specificity was 89.6% (SD 26.5%). Results were very similar for males and females (Table 1); in discriminating recordings from female participants, the mean sensitivity was 99.4% (SD 7.1%) and mean specificity was 85.7% (SD 34.1%), whereas for male participants, the mean sensitivity was 100% (SD 0%) and mean specificity was 88.9% (SD 31.6%). Stratification of data based on sex resulted in too few recordings to adequately fit the random forest classifier, which reduces the reliability of analysis and inference, particularly for *LRRK2*-PD (n = 7). Statistically significant differences were observed between *LRRK2*-PD and iPD voice features (see Figure 1 (showing clear separation plotting 2 salient features) and Supplementary Figure 1). Details regarding the most salient features are provided (Supplementary Table 3).

Accuracies were also computed using leave-one-subject-out (LOSO) CV [31]. Using LOSO CV, the mean sensitivity was 83.7% and mean specificity was 88.5% in discriminating *LRRK2*-PD from iPD (see Supplementary Table 2).

We performed additional analysis whereby non-manifesting *LRRK2* carriers and individuals with *LRRK2*-associated PD were treated as belonging to the same clinical group. This resulted in a larger sample of *LRRK2* carriers (*n*=27) which helped improve statistical power. The rationale of this analysis was to investigate if vocal deficits in *LRRK2* carriers (both manifesting and non-manifesting) were different from iPD (for details, see Supplementary analysis).

***Non-manifesting carriers (NMC) versus Related Non-carriers (RNC) and Controls***

Participants from the three groups were of similar age (Table 1). NMC had a higher UPDRSIII score compared to both healthy controls and RNC. Statistical analyses were performed using 39 NMC recordings (n = 20), 48 RNC recordings (n = 25), and 47 control recordings (n = 26). In discriminating NMC from RNC, the mean sensitivity was 74.9% (SD 24.0%) and mean specificity was 78.0% (SD 23.3%). Moreover, in discriminating NMC from unrelated healthy controls, the mean sensitivity was 75.7% (SD 24.3%) and mean specificity was 81.8% (SD 20.4%). Scatterplots of the most salient features for these pairwise comparisons do not allow readily visible identification of this discrimination (Figure 1). Compared to *LRRK2*-PD and iPD features, therefore, the voice features for NMC, RNC, and healthy controls were much less discriminatory.

**Discussion**

Our preliminary analyses found statistically significant differences between *LRRK2*-PD and iPD ($p<0.01$) in features extracted from sustained phonations (Figure 1 and Supplementary Figure 1). The differences in the features were less pronounced when non-manifesting carriers were compared with related non-carriers and healthy controls. Thus, voice could potentially be used as a non-invasive and inexpensive biomarker for identifying a *LRRK2* mutation in PD participants, but seems to be less promising as a potential marker for the prodromal phase of *LRRK2*-PD.

A limitation of this study is the small sample size, particularly for participants with *LRRK2*-PD. We investigated and verified that unique participant identity was not a confounder (see Supplementary analysis) [32]. *LRRK2*-PD participants were older than iPD participants and we cannot rule out the effect of presbyphonia as a potential confound [33]. However, including age as a covariate in the classification model did not result in improved classification accuracy, indicating that if presbyphonia exists in this cohort its effects may be negligible. Moreover, the predictive accuracy obtained using machine learning algorithms and multiple features do not lead to ready etiologically-relevant explanations for why voice impairment might be discriminatory in this context [34]. This hinders our ability to make inferences regarding underlying pathophysiological changes associated with an impaired voice in PD.

We find that statistical analysis of sustained phonations help discriminate *LRRK2*-PD and iPD. The findings of this study add to the growing evidence supporting clinical and pathological differences between *LRRK2*-PD and iPD, whereby differences in both motor and nonmotor features (including heart rate variability, tremor, gait, olfactory identification) have been previously reported [35-40]. To the best of our knowledge, this is the first proof-of-concept study that investigates potential vocal deficits in manifesting and non-manifesting *LRRK2*-carriers. These results warrant further investigation into the potential of using voice for the delineation of PD subtypes in larger cohorts.

**Acknowledgements**

We extend our sincere gratitude to all the participants who took part in this study and made this research possible.

**Funding:**

This study was funded by the Michael J Fox Foundation for Parkinson's Research.

TABLE 1. Descriptive statistics by group (A), comparison of clinical and demographic characteristics (B), and discrimination accuracy for pairwise comparisons (C).

| A. Descriptive statistics by group | n | Age Mean (SD) | % female | UPDRSIII Mean (SD) | Disease duration Mean (SD) |
|---|---|---|---|---|---|
| LRRK2-PD | 7 | 72.3 (13.1) | 57% | 21.9 (18.7) | 10.3 (11.8) |
| Idiopathic PD (iPD) | 17 | 63.4 (8.7) | 53% | 22.8 (10) | 5.4 (5.8) |
| Non-manifesting carriers (NMC) | 20 | 58.3 (14.5) | 45% | 4.1 (3.7) | n.a. |
| Related non-carriers (RNC) | 25 | 53.7 (16.3) | 64% | 2.7 (4.2) | n.a. |
| Healthy control (unrelated) | 26 | 54.0 (16.1) | 69% | 1.6 (1.8) | n.a. |
| B. Comparing descriptive statistics (p values) | | Age | | UPDRSIII | Disease duration |
| LRRK2-PD vs iPD | | 0.06 | | 0.55 | 0.21 |
| NMC vs RNC | | 0.33 | | ***0.04*** | - |
| NMC vs Healthy | | 0.36 | | ***0.01*** | - |
| C. Discrimination accuracy | | Sensitivity (%) Mean (SD) | | Specificity (%) Mean (SD) | |
| LRRK2-PD vs iPD (ALL) | | 95.4% (17.8%) | | 89.6% (26.5%) | |
| LRRK2-PD vs iPD (FEMALE) | | 99.4% (7.1%) | | 85.7% (34.1%) | |
| LRRK2-PD vs iPD (MALE) | | 100.0% (0%) | | 88.9% (31.6%) | |
| NMC vs RNC (ALL) | | 74.9% (24.0%) | | 78.0% (23.3%) | |
| NMC vs RNC (FEMALE) | | 80.2% (32.1%) | | 71.7% (36.5%) | |
| NMC vs RNC (MALE) | | 90.6% (22.6%) | | 85.1% (28.9%) | |
| NMC vs Healthy (ALL) | | 75.7% (24.3%) | | 81.8% (20.4%) | |
| NMC vs Healthy (FEMALE) | | 87.3% (26.3%) | | 84.6% (28.6%) | |
| NMC vs Healthy (MALE) | | 75.4% (34.8%) | | 82.1% (31.0%) | |

The table is presented in three different sections (A-C). Section **A** presents descriptive statistics for the five clinical groups (1. *LRRK2*-PD, 2. iPD, 3. NMC, 4. RNC, and, 5. Healthy controls). Section **B** compares the descriptive statistics. Age was analyzed using an unpaired t-test, whereas disease duration and UPDRSIII total were compared using Mann–Whitney U test. *p* values <0.05 are highlighted in the bold italic text. Section **C** presents the out-of-sample discrimination accuracy for the three priority pairwise comparisons (1. *LRRK2*-PD vs iPD, 2. NMC vs RNC, and, 3. NMC vs Healthy) using a 10-fold cross-validation (CV) scheme (with 100 repetitions), employing only the 10 most salient voice features. The scheme involved repetitive splitting of the data into a training set (90% of the total observations) and a validation set (remaining 10% of the observations). The mean sensitivity and mean specificity across different CV repetitions are presented (along with standard deviation in brackets). The data was balanced to account for differences in group sample size. Accuracies were computed using all recordings, and separately for subgroup analysis using data stratified by sex. Abbreviations used: iPD, idiopathic Parkinson's disease; *LRRK2*-PD, *LRRK2*-associated Parkinson's disease; NMC, non-manifesting carriers; RNC, related non-carriers; SD, standard deviation; UPDRSIII, Movement Disorders Society Unified Parkinson's Disease Rating Scale part 3; n.a., not applicable.

FIGURE 1 TITLE. Scatterplots and boxplots of salient features for the three pairwise comparisons: *LRRK2*-PD versus idiopathic PD (iPD) (Panels A and B), non-manifesting carriers (NMC) versus related non-carriers (RNC) (Panels C and D), and NMC versus healthy controls (Panels E and F).

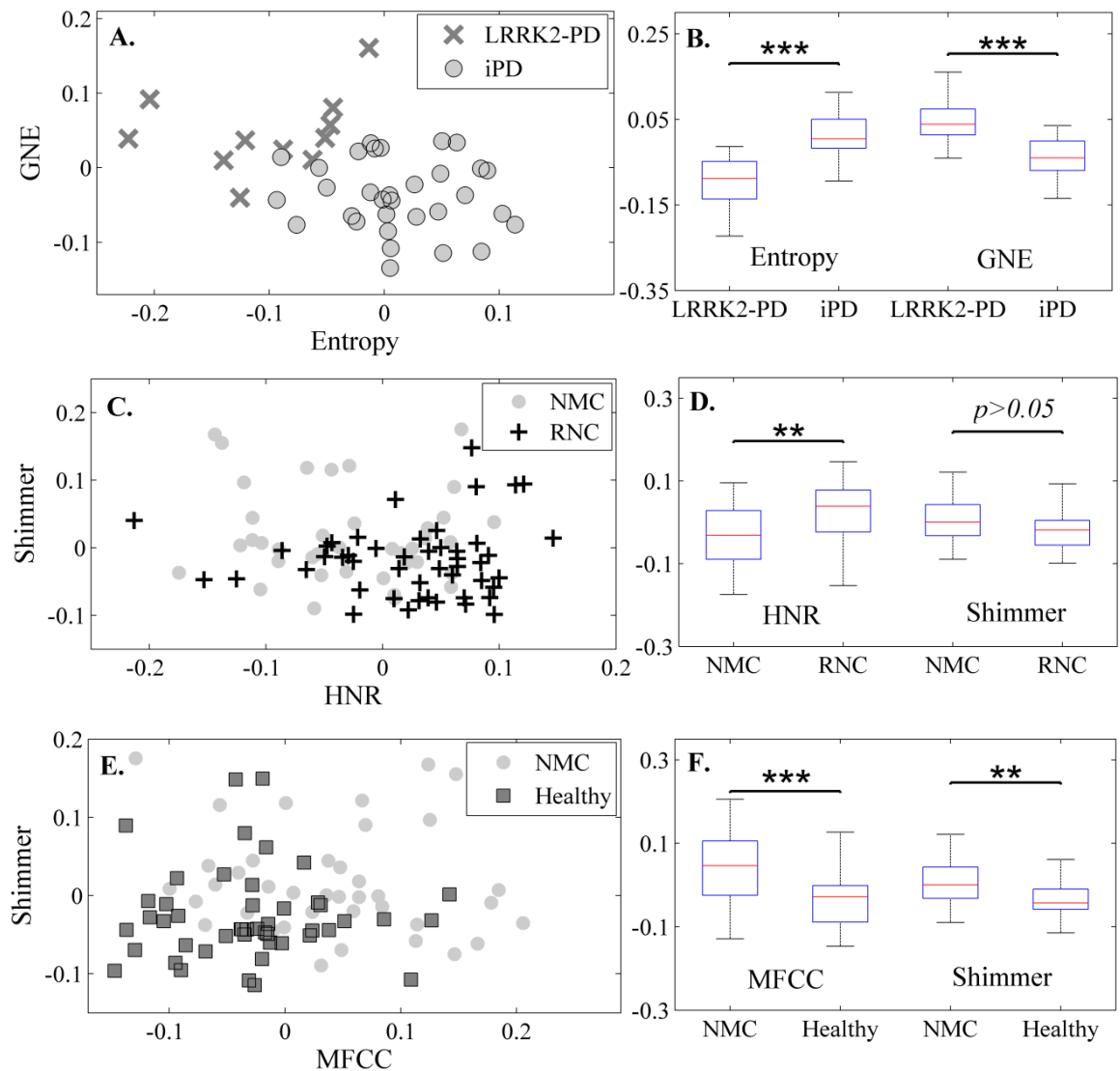

Panel A shows two highly discriminatory features that help differentiate *LRRK2*-PD from iPD. In Panel A, we plot Entropy (entropy computed after wavelet decomposition, quantifies extent of randomness in a signal) and Glottis to Noise Excitation (GNE, degree of signal strength versus noise resulting from incomplete vocal fold closure), both features were significantly different ($p<0.001$, denoted by ***) (Panel B). Panel C plots two salient features that help discriminate NMC from RNC. In Panel C, we plot Harmonic to Noise Ratio (HNR, signal to noise ratio) and median shimmer (roughness in voice). HNR between the two groups was significantly different ($p<0.01$, denoted by **), whereas shimmer between NMC and RNC was similar ($p>0.05$) (Panel D), which indicates that the two cohorts are less different (as reflected in the discrimination accuracies reported in Table 1). Panel E shows two salient features that discriminate NMC from healthy controls. In Panel E, we plot Mel Frequency Cepstral Coefficient (MFCC, quantifies vocal fold dynamics depending on properties of the articulators) and median shimmer. Panel F shows that MFCC and shimmer were significantly different between the two groups. Salient features were identified separately for each pairwise comparison, using five different feature selection algorithms. The above plots were generated using all usable voice recordings. *p* values reported above were computed using the nonparametric two-sided Kolmogorov-Smirnov (KS) test.

Supplementary material for **'Investigating Voice as a Biomarker for leucine-rich repeat kinase 2-Associated Parkinson's Disease'**

**Supplementary Analysis**

*Feature extraction*

For each recording, we extracted 292 summary measures (also referred to as *features*). These features can broadly be characterized as: (1) *Descriptive features*: statistical characteristics of voice. (2) *Vocal fold vibration-based features*: variation in speech frequency and amplitude. (3) *Cepstral coefficients*: subtle changes in the placement of the articulators (mouth, tongue, teeth and lips). (4) *Aeroacoustics-based features*: degree of turbulent noise in speech due to incomplete vocal fold closure. (5) *Wavelet features*: extended time-frequency domain properties. Details regarding these feature categories are provided in Supplementary Table 1.

*Feature selection*

For each pairwise comparison, salient features were identified using the following 5 feature selection algorithms that help enhance the explanatory power of the analysis by removing redundant and less informative features: (1) Minimum redundancy maximum relevance (mRMR) [27], (2) Gram-Schmidt orthogonalization (GSO) [25], (3) RELIEF [26], (4) Local learning-based feature selection (LLBFS) [28], and (5) Least absolute shrinkage and selection operator (LASSO) [29].

*Validation*

In this study, we used two model validation schemes: (1) 10-fold cross-validation: this scheme involves randomly splitting the data into two non-overlapping parts, the first part of the data (comprising 90% of the recordings) are used to train the model (i.e., learn the underlying differences in patterns of voice-based features for each pairwise comparison), while the remaining 10% of the recordings are used for validation (i.e., evaluate the accuracy of model predictions). This process of randomized selection of training and validation sets was repeated multiple times, and the discrimination accuracies (quantified using sensitivity and specificity) were calculated on each repetition. (2) Leave-one-subject-out: this scheme involves splitting the data such that all recordings from only one participant are used for model validation, while all remaining recordings are used for training. This process is repeated multiple times. Note that for both schemes, the accuracy of model predictions are computed using only the validation set. Basically, the model is blinded to the validation set during the training process, which helps gauge generalizability of the model to previously unseen similar datasets.

In this study, we analysed all available/suitable voice recordings leading to a mismatch in the five group sizes (Table 1). Whilst there are fewer *LRRK2*-associated PD participants compared to iPD, it should be noted that we used a 'balanced cross-validation scheme' that results in an equal number of samples across different classes for each pairwise comparison. This scheme helps mitigate the issues associated with imbalanced datasets/differences in group sizes.

***LRRP2-PD vs idiopathic PD* (excluding 1 *LRRK2*-PD participant)**

*LRRP2*-PD participants had longer mean disease duration (n = 7, mean = 10.3 years, SD = 11.8 years) compared to participants with iPD (n = 17, mean = 5.4 years, SD = 5.8 years). One *LRRK2*-PD participant had a disease duration of 36 years, and without this participant, the mean disease duration for the remaining 6 *LRRK2*-PD participants was 6 years (SD = 3.7 years), which is similar to the mean disease duration for iPD participants (5.4 years). Statistical analyses were thus performed separately by excluding this *LRRK2*-PD participant (disease duration 36 years, female, age 85 years, one voice recording available). The rationale for this analysis was to investigate disease duration as a potential confounding factor in discriminating *LRRK2*-PD versus iPD. Using 10 voice recordings collected from the remaining 6 *LRRK2*-PD participants, we recomputed the accuracy in discriminating *LRRK2*-PD versus iPD.

Using only the 10 top-ranked features for this pairwise comparison, the mean sensitivity and mean specificity was 97.0% (SD 15.5%) and 87.2% (SD 31.1%) in discriminating *LRRK2*-PD from iPD using all recordings, 99.0% (SD 10.0%) and 82.4% (SD 38.1%) in discriminating *LRRK2*-PD from iPD using only female recordings, and 100% (SD 0%) and 88.9% (SD 31.6%) in discriminating *LRRK2*-PD from iPD using only male recordings. These accuracies were obtained using 10-fold cross-validation. Note that the excluded *LRRK2*-PD recording was collected from a female participant; hence the accuracy in discriminating *LRRK2*-PD versus iPD using only male recordings was the same as those reported in Table 1. For leave-one-subject-out cross-validation, mean sensitivity and mean specificity were 88.5% (SD 7.8%) and 81.3% (SD 12.3%) respectively, in discriminating *LRRK2*-PD from iPD using all recordings. The sample size was too small to draw any reliable inference based on discrimination accuracies for subgroup analysis stratified by sex. These sensitivity and specificity values (obtained using recordings from n = 6 *LRRK2*-PD participants) are in close agreement with the discrimination accuracies obtained using all available recordings for *LRRK2*-PD (n = 7), as reported in Table 1 and Supplementary Table 2. Encouragingly, for all pairwise comparisons reported above, the sensitivity and specificity results differed statistically significantly from comparable results obtained from completely randomized predictions about which participants had a *LRRK2* mutation or were iPD (these predictions are akin to outcomes of an unbiased coin flip and are based on chance alone). Moreover, the differences in *LRRK2*-PD and iPD voice recordings is also evident from the scatterplot of salient features that shows two distinct clusters, as presented in Supplementary Figure 4.

*Detecting and characterising identity confounding*

Digital recordings of voice and other sensor data from individuals can capture properties of these data which are unique to particular individuals, in the following way. For example, in voice, the combination of various vocal features such as vocal pitch and spectral envelope may occupy an approximately unique region in feature space, distinct from all other individuals in the study. This uniqueness can interact with highly nonlinear classifiers to produce 'identity confounding' whereby the classifier finds a relationship between the individual and their specific clinical grouping, rather than a relationship between clinical symptoms and clinical grouping. This inadvertent relationship can confound predictions, which means that it is necessary to quantify the extent to which the classifier is making predictions which would generalize to individuals not in the study. To quantify this potential confound, we counted the number of observations and individuals per unique predicted value in the classification tree (across 500 trees used); whereby each tree was built using a bootstrap sample of the training data. Averaged over all cross-validation repetitions (10-fold with 100 repetitions), we found that the average number of observations per unique predicted value and corresponding number of participants were 8.2 and 4.5, respectively, for *LRRK2*-PD versus iPD. The average number of participants in the training set (after balancing the data and bootstrapping) in each tree for the above pairwise comparison was 10.9. While training the model, the unique predicted values thus had observations from around 41% of the individuals that were in the training set (4.5 individuals on average), thus indicating that this form of identity confounding is unlikely to be a significant factor in the results presented in this study.

### *Prodromal versus Nonprodromal*

We investigated if two non-manifesting carriers classified as being in the prodromal state were more similar to their non-prodromal counterparts compared to participants with *LRRK2*-PD. One of the two participants meeting prodromal criteria was more similar to the manifesting LRRK2 carriers on the basis of the two most salient voice features, (Supplementary Figure 3). However, the sample size was too small to draw any reliable inferences.

### *LRRK2 carriers vs idiopathic PD*

For the *LRRK2* carrier group, we analyzed 50 recordings, collected from 27 individuals (mean age: 61.9 years (SD 15.3); % female: 48.2%; mean UPDRSIII: 8.7 (SD 12.4)), whereas for the iPD group, we analyzed 32 recordings, collected from 17 individuals (mean age: 63.4 years (SD 8.7); % female: 53.0%; mean UPDRSIII: 22.8 (SD 10.0)). Age for the *LRRK2* carrier and iPD groups were similar at 5% significance level (unpaired t-test). As expected, UPDRS III for the *LRRK2* carrier group was significantly lower compared to the iPD group (Mann–Whitney U test). In distinguishing *LRRK2* carriers and iPD, using the 10 most salient features, the mean sensitivity was 74.8% (SD 26.5%) and mean specificity was 83.0% (SD 22.3%). We used 10-fold CV with 100 repetitions. The differences in the features for *LRRK2*-carriers versus iPD were thus less pronounced, compared to the case when features for *LRRK2*-PD were compared against iPD. Further investigations are using larger cohorts are needed to investigate if non-manifesting and manifesting *LRRK2* carriers can be treated as belonging to the same clinical group.

## LIST OF SUPPLEMENTARY TABLES

**SUPPLEMENTARY TABLE 1.** Brief description of features extracted from the voice recordings.

**SUPPLEMENTARY TABLE 2.** Discrimination accuracy for the leave-one-subject-out (LOSO) cross-validation (CV) scheme for the three pairwise comparisons: *LRRK2*-associated Parkinson's disease (*LRRK2*-PD) versus idiopathic PD (iPD), non-manifesting *LRRK2* mutation carriers (NMC) versus related non-carriers (RNC), and NMC versus healthy controls, computed using a machine learning algorithm (random forest) and a naïve discrimination benchmark (randomized predictions).

**SUPPLEMENTARY TABLE 3.** List of 10 salient features selected for three pairwise comparisons.

## LIST OF SUPPLEMENTARY FIGURES

**SUPPLEMENTARY FIGURE 1.** Scatterplots and boxplots of salient features for the pairwise comparison: *LRRK2*-associated Parkinson's disease (*LRRK2*-PD) versus idiopathic PD (iPD).

**SUPPLEMENTARY FIGURE 2.** Discrimination accuracies as a function of the number of salient features used in the machine learning discrimination analysis, for the three pairwise comparisons: *LRRK2*-associated Parkinson's disease (*LRRK2*-PD) versus idiopathic PD (iPD), non-manifesting *LRRK2* mutation carriers (NMC) versus related non-carriers (RNC), and NMC versus healthy controls.

**SUPPLEMENTARY FIGURE 3.** Scatterplot of two most salient features for the pairwise comparison: *LRRK2*-associated Parkinson's disease (*LRRK2*-PD) versus non-manifesting *LRRK2* mutation carriers (NMC), plotted along with voice features extracted from prodromal participants.

**SUPPLEMENTARY FIGURE 4.** Scatterplot of two most salient features for the pairwise comparison: *LRRK2*-associated Parkinson's disease (*LRRK2*-PD) versus idiopathic PD (iPD), plotted along with voice features from the excluded *LRRK2*-PD participant.

SUPPLEMENTARY TABLE 1: Brief description of five categories of features extracted from the voice recordings.

| Category | Brief description |
| --- | --- |
| *Category 1: Descriptive features:* | |
| Mean, median, standard deviation, skewness, interquartile range etc. | Quantifies statistical characteristics of the voice signal |
| *Category 2: Vocal fold vibration-based features:* | |
| Jitter | Quantifies the instabilities of the oscillating pattern of the vocal folds by measuring cycle-to-cycle changes in the fundamental frequency (measure of roughness in voice) |
| Shimmer | Quantifies the instabilities of the oscillating pattern of the vocal folds by measuring cycle-to-cycle changes in the amplitude (measure of roughness in voice) |
| Teager-Kaiser Energy Operator (TKEO) | Measures the instantaneous changes in voice energy (takes into account both amplitude and frequency) |
| *Category 3: Cepstral coefficients based features:* | |
| Mel Frequency Cepstral Coefficients (MFCCs) | Computes the contribution of the energy of the speech signal at each frequency band (are aimed at detecting subtle changes in the motion of the articulators) |
| *Category 4: Aeroacoustics, aperiodicity, and frequency based features:* | |
| Recurrence Period Density Entropy (RPDE) | Quantifies any ambiguity in fundamental pitch (RPDE is zero for perfectly periodic signals and one for purely stochastic signals). Higher RPDE has been associated with voice impairment |
| Detrended Fluctuation Analysis (DFA) | Characterizes the changing detail of aero-acoustic breath noise |
| Pitch Period Entropy (PPE) | Measures the impaired control of stable pitch, a property common in PD |
| Harmonics-to-Noise Ratio (HNR) | Quantifies noise in the speech signal, caused mainly due to incomplete vocal fold closure |
| Glottal to Noise Excitation (GNE) ratio | Quantifies the extent of noise in speech using linear and nonlinear energy measures |
| Vocal Fold Excitation Ratios (VFER) | Quantifies the extent of noise in speech using energy (linear and nonlinear) and entropy-based measures |
| Perturbation Quotient (PQ) | Quantifies variations in speech signal |
| Glottis Quotient (GQ) | Quantifies properties of the vocal folds (when glottis is open and closed) |
| F0 contour features | Measures based the summary statistics of the fundamental frequency |
| *Category 5: Wavelet-based features:* | |
| Wavelet related measures | Variants of above-discussed summary measures applied to wavelet coefficients of the speech signal |

**SUPPLEMENTARY TABLE 2.** Discrimination accuracy for the leave-one-subject-out (LOSO) cross-validation (CV) scheme for the three pairwise comparisons: *LRRK2*-associated Parkinson's disease (*LRRK2*-PD) versus idiopathic PD (iPD), non-manifesting *LRRK2* mutation carriers (NMC) versus related non-carriers (RNC), and NMC versus healthy controls, computed using a machine learning algorithm (random forest) and a naïve benchmark (randomized predictions).

| *Discrimination accuracy* | Sensitivity (%) Mean (SD) | Specificity (%) Mean (SD) |
|---|---|---|
| **LRRK2-PD vs iPD *(ALL)*** | | |
| Random forest | 83.7% (7.1%) | 88.5% (8.4%) |
| Randomized predictions | 51.6% (20.2%) | 47.0% (18.3%) |
| | | |
| **NMC vs RNC *(ALL)*** | | |
| Random forest | 67.3% (3.2%) | 69.8% (4.9%) |
| Randomized predictions | 48.6% (13.1%) | 47.3% (11.1%) |
| | | |
| **NMC vs Healthy *(ALL)*** | | |
| Random forest | 72.4% (3.2%) | 69.9% (5.0%) |
| Randomized predictions | 50.4% (10.0%) | 48.9% (11.6%) |

The above sensitivity and specificity values were computed separately for each of the three priority pairwise comparisons (1. *LRRK2*-PD vs iPD, 2. NMC vs RNC, and, 3. NMC vs Healthy) using a leave-one-subject-out (LOSO) cross-validation (CV) scheme, employing 10 most salient voice features. Validation scheme involved repetitive splitting of the data such that at a given CV iteration, all voice tests from only one randomly selected participant were employed for model validation, while voice tests from all remaining participants were used for training. We used LOSO CV scheme with 100 repetitions. The data was balanced to account for differences in number of participants in each clinical group. Accuracies are reported for a machine learning classifier (random forest) and a naïve benchmark based on randomized predictions (expected accuracy around 50%), using all available voice tests from the five clinical groups (1. *LRRK2*-PD, 2. iPD, 3. NMC, 4. RNC, and, 5. Healthy controls). The sensitivity and specificity values were presented in percentage (%) as mean (and standard deviation, in brackets), whereby the standard deviation denotes the variability in the accuracy across multiple CV repetitions. The rankings of the most salient features were obtained separately for each of the three pairwise comparisons, using a majority voting scheme (using 5 feature selection algorithms). Abbreviations used: iPD, idiopathic Parkinson's disease; *LRRK2*-PD, *LRRK2*-associated Parkinson's disease; NMC, non-manifesting carriers; RNC, related non-carriers; SD, standard deviation.

SUPPLEMENTARY TABLE 3: List of 10 salient features selected for the three pairwise comparisons.

| Feature Name | Brief description |
|---|---|
| *Comparison 1*: LRRK2-associated PD vs iPD | |
| Skewness | Quantifies asymmetry of the distribution |
| det_entropy_log_6_coef | Wavelet log-entropy of the $6^{th}$ detail coefficient of F0, quantifies subtle changes in the details of F0 fluctuations |
| GNE-SEO | Glottal to Noise Excitation Squared Energy Operator, quantifies excessive noise and turbulence in the voice |
| VFER-LF-TKEO | Vocal Fold Excitation Ratio, quantifies incomplete vocal fold closure which creates vortices and inconsistencies across frequency bands in terms of energy |
| prctile50TKEO_A0 | Median of the Teager-Kaiser Energy of amplitude |
| app_det_TKEO_mean_4_coef | Mean Teager-Kaiser Energy of the $4^{th}$ wavelet decomposiition coefficient decomposing F0, quantifies subtle changes in the energy of F0 |
| medMFCC3 | Median value of the $3^{rd}$ MFCC coefficient, quantifies envelope structure fluctuations |
| app_entropy_log_8_coef | Entropy of the $8^{th}$ approximation wavelet decomposition coefficient, quantifies changes in F0 |
| det_entropy_shannon_6_coef | Shannon entropy of the $6^{th}$ detail wavelet decomposition coefficient, quantifies changes in F0 |
| Q1 | $25^{th}$ quartile |
| *Comparison 2*: Non-manifesting carriers (NMC) versus related controls | |
| det_LT_TKEO_mean_4_coef | Mean Teager-Kaiser Energy of the $4^{th}$ detail wavelet decomposition coefficient, quantifies changes in F0 |
| HNR(1) | Harmonics to Noise Ratio, quantifies signal to noise, i.e. the extent of vocal noise using standard autocorrelation |
| Mean(A0) | Mean amplitude |
| medShimmer | Quantifies amplitude perturbations |
| PQ11.class_Schoentgen | Amplitude perturbation using a 11-sample window |
| muDiffMFCC5 | $5^{th}$ MFCC |
| medMFCC10 | Median of $10^{th}$ MFCC, quantifies mostly higher harmonic components in the signal |
| medJitter | Quantifies frequency perturbations |
| mode_F0 | Dominating F0 value |
| det_LT_TKEO_mean_8_coef | Mean Teager-Kaiser Energy of the $8^{th}$ detail wavelet decomposition coefficient |
| *Comparison 3*: Non-manifesting carriers (NMC) versus unrelated controls | |
| muDiffMFCC13 | $13^{th}$ MFCC |
| muDiffMFCC8 | Quantifies mostly higher harmonic components in the signal |
| medShimmer | Quantifies amplitude perturbations |
| Ed2_8_coef | Wavelet energy of the $8^{th}$ wavelet coefficient |
| PQ11.class_Schoentgen | Amplitude perturbation using a 11-sample window |
| medMFCC10 | Quantifies mostly higher harmonic components in the signal |
| Ed2_7_coef | Wavelet energy of the $7^{th}$ wavelet coefficient |
| det_entropy_log_6_coef | Wavelet log energy the $6^{th}$ detail wavelet decomposition coefficient, quantifies changes in F0 |
| P0 | Perturbation quotient (zeroth order) |
| HNR(1) | Harmonics to Noise Ratio, quantifies signal to noise, i.e. the extent of vocal noise using standard autocorrelation |

**SUPPLEMENTARY FIGURE 1.** Scatterplots and boxplots of salient features for the pairwise comparison: *LRRK2*-associated Parkinson's disease (*LRRK2*-PD) versus idiopathic PD (iPD).

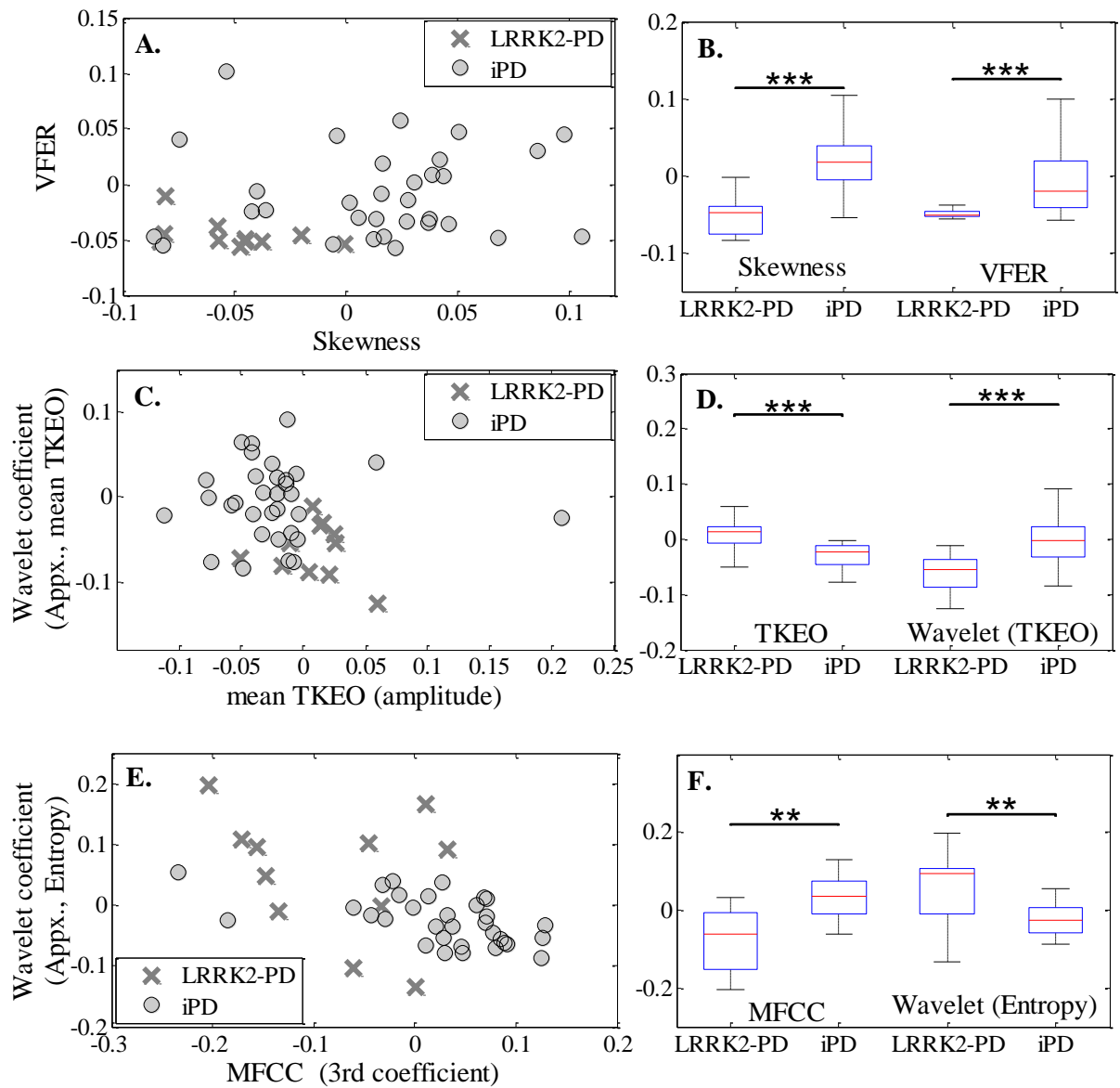

Panel A plots two salient features, Skewness (amplitude) and Vocal Fold Excitation Ratio (VFER, the degree of signal strength over noise resulting from incomplete vocal fold closure), both skewness and VFER were significantly different between the two groups, ($p<0.001$, denoted by ***) (Panel B). Panel C plots mean Teager Kaiser Energy Operator (TKEO, quantifies instantaneous changes in voice energy) and a Wavelet coefficient (based on TKEO), while Panel D shows that these features were significantly different between *LRRK2*-PD and iPD. Panel E plots the Mel Frequency Cepstral Coefficients (MFCC, quantifies vocal fold dynamics taking into account the properties of the articulators) and Entropy (entropy computed after wavelet decomposition, computes the extent of randomness in a signal), while Panel F shows that these features were statistically significantly different ($p<0.01$). Features with high discriminatory power were identified using five different feature selection algorithms. The above plots were generated using all voice recordings collected from participants with *LRRK2*-PD and iPD. *p* values reported above were computed using the nonparametric two-sided Kolmogorov-Smirnov (KS) test.

**SUPPLEMENTARY FIGURE 2.** Discrimination accuracies as a function of the number of salient features used in the machine learning discrimination analysis, for the three pairwise comparisons: *LRRK2*-associated Parkinson's disease (*LRRK2*-PD) versus idiopathic PD (iPD) (Panels A and B), non-manifesting *LRRK2* mutation carriers (NMC) versus related non-carriers (RNC) (Panels C and D), and NMC versus healthy controls (Panels E and F).

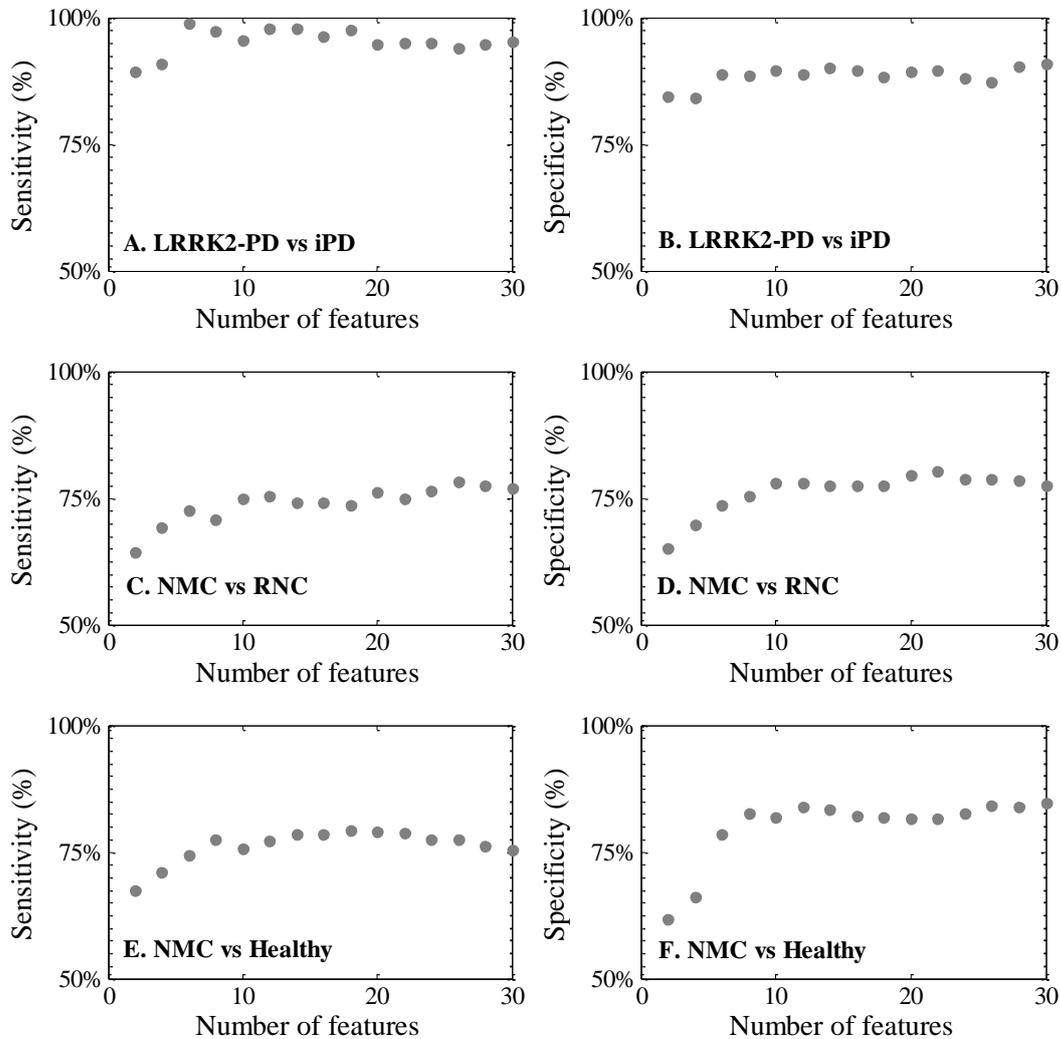

The above accuracies were computed using all available voice recordings from the five clinical groups (1. *LRRK2*-PD, 2. iPD, 3. NMC, 4. RNC, and, 5. Healthy controls), using 10-fold cross-validation (100 repetitions). The rankings of the most salient features were obtained using a majority voting scheme (using 5 feature selection algorithms). The feature rankings were obtained separately for each of the above 3 pairwise comparisons (1. *LRRK2*-PD vs iPD, 2. NMC vs RNC, and, 3. NMC vs Healthy). Features were added into the machine learning classifier (random forest) in increments of 2 (starting from 2, and going up to 30), whereby higher ranked features were added first. The whole process of training and validation was repeated each time two new features were included. Mean sensitivity and specificity values are denoted as grey circles and reported in percentage (%).

**SUPPLEMENTARY FIGURE 3.** Scatterplot of two most salient features that help discriminate *LRRK2*-associated Parkinson's disease (*LRRK2*-PD) versus non-manifesting *LRRK2* mutation carriers (NMC), plotted along with features from prodromal participants.

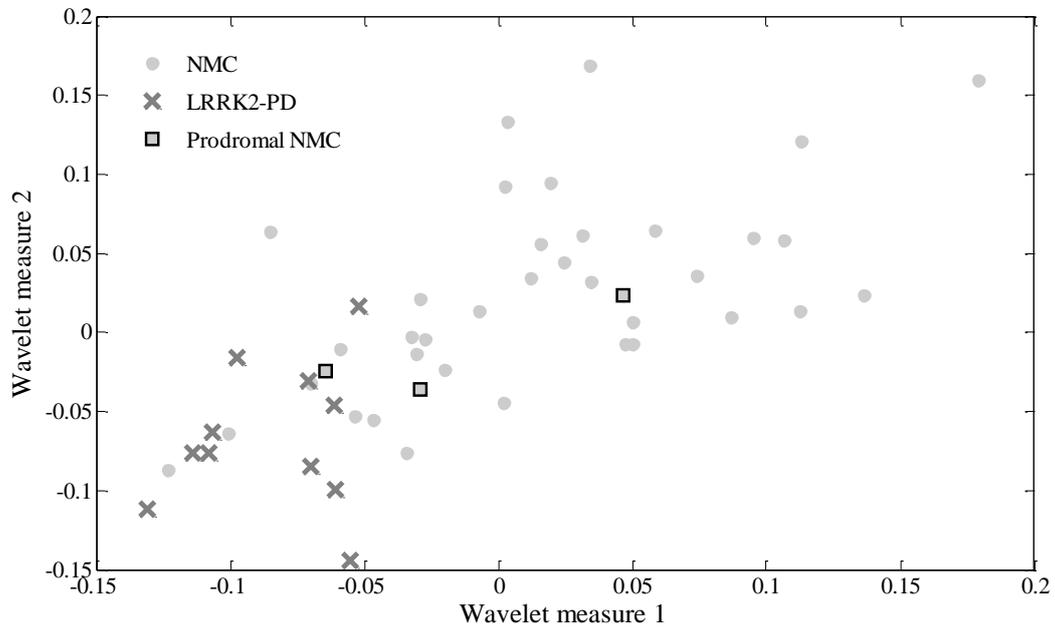

The feature rankings were obtained separately for the above pairwise comparison: *LRRK2*-PD versus NMC. We analysed three voice recordings from two prodromal participants (denoted as a grey square).

**SUPPLEMENTARY FIGURE 4.** Scatterplot of two most salient features that help discriminate *LRRK2*-associated Parkinson's disease (*LRRK2*-PD) versus idiopathic PD (iPD), plotted along with voice features from the excluded *LRRK2*-PD participant.

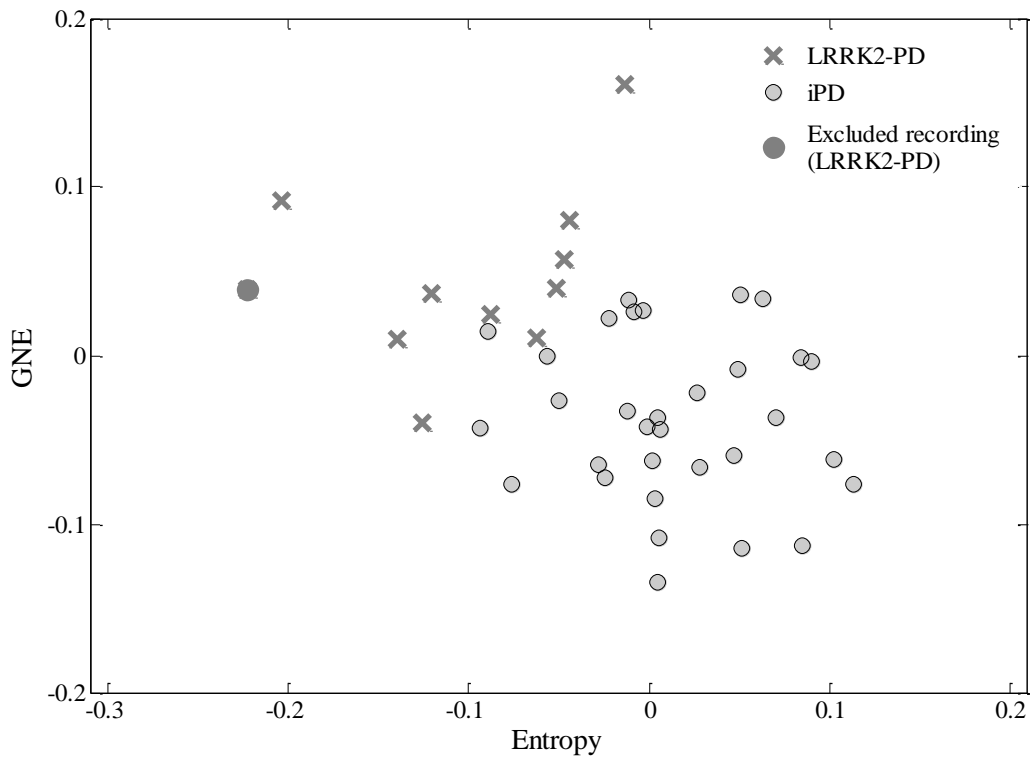

We analysed 10 voice recordings collected from six *LRRK2*-PD participants (denoted as grey crosses) and 32 voice recordings obtained from seventeen iPD participants (denoted as light grey circles). Analysis excluded one *LRRK2*-PD participant who had disease duration of 36 years, note that we only had one decent quality voice recording for this participant (denoted as a dark grey circle).